\documentclass[secnumarabic,twocolumn,graphics,floatfix,superscriptaddress,tightenlines,aps,prb]{revtex4-1}

\usepackage{epsf}
\usepackage{bm}

\begin{document}

\title{Complexes of dipolar excitons in layered quasi-two-dimensional nanostructures}

\author{Igor~V.~Bondarev}
\affiliation{Department of Mathematics \& Physics, North Carolina Central University, Durham, NC 27707, USA}
\author{Maria~R.~Vladimirova}
\affiliation{Laboratoire Charles Coulomb, UMR 5221 CNRS-Universit\'{e} de Montpellier, F-34095, Montpellier, France}

\begin{abstract}
We discuss neutral and charged complexes (biexciton and trion) formed by indirect excitons in layered quasi-two-dimensional semiconductor heterostructures. Indirect excitons --- long-lived neutral Coulomb-bound pairs of electrons and holes of different layers --- have been known for semiconductor coupled quantum wells and are recently reported for van der Waals heterostructures such as bilayer graphene and transition metal dichalcogenides. Using the configuration space approach, we derive the analytical expressions for the trion and biexciton binding energies as functions of the interlayer distance. The method captures essential kinematics of complex formation to reveal significant binding energies, up to a few tens of meV for typical interlayer distances $\sim\!3-5$~\AA, with the trion binding energy always being greater than that of the biexciton. Our results can contribute to the understanding of more complex many-body phenomena such as exciton Bose-Einstein condensation and Wigner-like electron-hole crystallization in layered semiconductor heterostructures.
\end{abstract}

\pacs{Valid PACS appear here}% PACS, the Physics and Astronomy

\maketitle

%\tableofcontents

\section{\label{sec:Intro} Introduction}

Assemblies of particles featuring permanent electric dipole moments are interesting due to their inherent anisotropic many-body effects resulted from the interparticle interaction anisotropy in the assembly. Dipolar many-body systems have been realized both in atomic physics and in solid state physics.\cite{Giovanazzi,Doyle2004,Ni2010} For layered solid state heterostructures, in particular, dipolar (or indirect) excitons --- Coulomb-bound electron-hole pairs in coupled semiconductor quantum wells (CQWs) where the electrons and holes are confined to \emph{different} quasimonolayers of one semiconductor separated by the quasimonolayer of another semiconductor --- have been known for several decades.\cite{LozovikYudson,Fukuzawa,LozovikBerman,Berman,ZimmermanPRB2008,Gorbunov2008,Larionov2008,HighNature,Oliveira2008,Voros06,Voros2009,Liu2017,Kirsanske2016,Shields97,Govorov13,
Kyriienko2012,Schinner2013,Rapaport,Anankine2017,Vina,Kowalik,Sivalertporn2012,Vishnevsky2013,Violante,Butov2016,Butov2002,Butov2004,Andreakou,Dietl2017,GRASSELLI2017,Wilkes2017,Butov2016,
Butov2017,Combescot2017} Indirect excitons in such systems can be coupled to the light modes of properly designed photonic microcavities to form dipolar exciton-polaritons (dipolaritons), offering control over quantum phenomena such as electromagnetically induced transparency, room-temperature Bose-Einstein condensation (BEC), and adiabatic photon-to-electron transfer.\cite{Szymanska2012,Cristofolini2012,Berman08,Berman16,Yamamoto14}

Very recently, indirect excitons have been observed in double bilayer graphene systems\cite{Li2017} and in layered qiasi-two-dimensional (quasi-2D) transition metal dichalcagenide (TMD) heterostructures.\cite{Fogler14,Calman2017,Ross2017,Rivera2015,Baranowski2017} Here, the electron and hole reside on neighboring monolayers, with their wave function overlap and associated exciton recombination rate greatly reduced. Such indirect excitons have long lifetimes and so are able to completely thermalize after the excitation, to reveal a variety of fundamental collective many-particle effects of equilibrium quantum statistics, including BEC, superfluidity, and Wigner crystallization earlier predicted theoretically but only observed experimentally in part by now. Recent reviews on physics and applications of cold dipolar excitons can be found in Refs.~\onlinecite{Butov2016,Combescot2017,Butov2017}.

In most of solid state semiconductor nanostructures excitons (direct, or indirect) coexist with free charge carriers.\cite{Anankine2017,Li2017,Butov2002,Butov2004,Butov2016,Butov2017,Fogler14,Calman2017,Combescot2017,Szymanska2012,Cristofolini2012,Larionov2008,Andreakou,Yamamoto14,
Berman,Rivera2015,Ross2017,Kim2016,Bellus2015,Baranowski2017} At high carrier densities, strong charge screening greatly reduces the probability of the bound electron-hole state formation.\cite{Zimmermann,HaugKoch} At intermediate career densities, the scattering of formed excitons by free carriers and lattice vibrations affects the collective properties of excitonic systems through their energy, momentum, phase, electrical and spin polarization relaxation,\cite{Andreakou,Berman,Larionov2008,Anankine2017,Manca2017,Wang17,BondarevNL17} resulting in the spectral broadening and temperature lineshape variations of the exciton resonances in the optical excitation spectra.\cite{Cadiz17,Cassabois17,BondarevAPL16,Moody2015,Bondarev2003} In the process of inelastic scattering, an exciton can capture an extra charge to form a \emph{charged} bound three-particle complex --- the trion.\cite{Kheng1993,BarJoseph2005,ButovJETP,Mak2013,Finley17} Such states were first predicted by Lampert for bulk semiconductors.\cite{Lampert1958} Similarly, when the density of excitons is high enough, their inelastic scattering can result in the formation of the \emph{neutral} complex of two excitons, the biexciton.\cite{Crottini,Miller1982,Plechinger2015,Birkedal,Forchel} Formation of biexcitons and trions, though not detectable in bulk materials at room temperature, plays a significant role in quantum confined systems of reduced dimensionality such as quantum wells,\cite{Birkedal,Miller1982,Plechinger2015,Bracker,Schuetz} nanowires,\cite{Forchel,Crottini} nanotubes,\cite{Matsunaga11,Santos11,Colombier12,Yuma13} and quantum dots.\cite{Woggon,JonFbiexc,JonFtrion} Biexciton and trion excitations open up routes for controllable nonlinear optics and spinoptronics applications, respectively. The trion, in particular, has both net charge and spin, and therefore can be controlled by electrical gates while being used for optical spin manipulation, or to investigate correlated carrier dynamics in low-dimensional materials.

\begin{figure}[t]
\epsfxsize=8.5cm\centering{\epsfbox{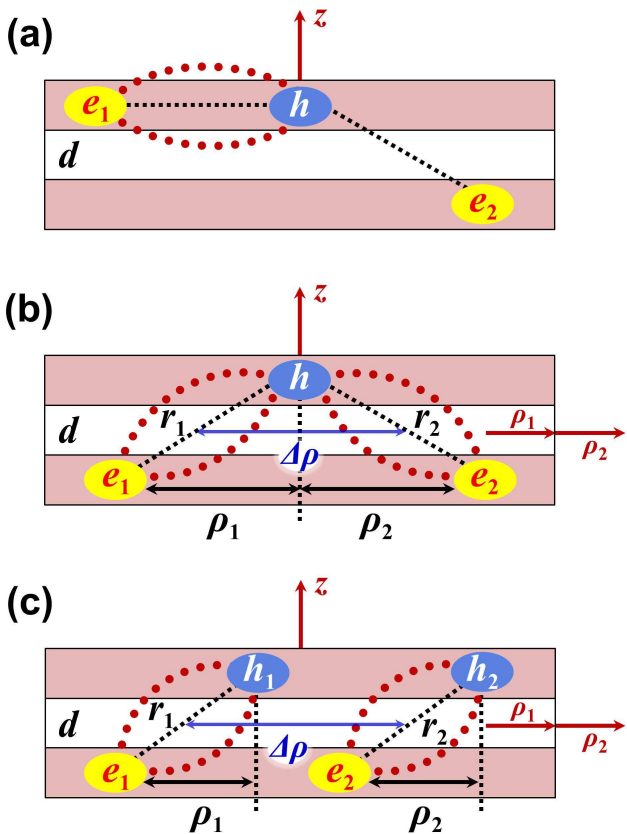}}\caption{(Color online) Sketch of the trion complex formed by an electron and a direct exciton (a) and that formed by an electron and an indirect exciton (b) in a bilayered structure with the interlayer distance $d$. In (a), exciton configurations ($h-e_1$) and ($h-e_2$) are inequivalent. In (b), they are equivalent (and can be viewed as two indirect excitons sharing the same hole to form a negatively charged trion state) and so the configuration space method can be used to evaluate the lowest ground state binding energy of the trion complex. (c)~Sketch of the biexciton complex of two indirect excitons.}\label{fig1}
\end{figure}

A rigorous theoretical description of compound quantum exciton systems such as biexciton and trion is a challenging problem. For conventional excitons (direct, with zero permanent dipole moment) in semiconductor nano\-structures of reduced dimensionality, a number of theoretical treatments can be found in the literature, predominantly relying on brute force computational techniques for the most part.\cite{Singh,Thilagam,Filinov2003,Lozovik,Pedersen05,Drummond05,Sergeev2005,Sidor,Kammerlander07,Zimmer08,Fogler08,
Bondarev11PRB,Kezer11,Ronnow11,Watanabe12,Ronnow12,Bondarev14PRB,Kezer14,Velizh15,Kezer17,Bondarev2016} Here, we consider a special case of the trion and biexciton states that are formed by the \emph{indirect} (dipolar) excitons in layered quasi-2D semiconductor heterostructures, whose binding energies we derive analytically as functions of the interlayer separation distance and other parameters of the system. Figure~\ref{fig1}~(a) and (b) shows the difference between the trion states formed by the direct and indirect excitons, respectively. The trion we study is the charged \emph{indirect} exciton --- a charged three-particle Coulomb bound complex formed by a pair of electrons (holes) and a hole (electron) in which two like charge particles are confined to the same layer and the (third) opposite charge particle is confined to another layer separated by the interlayer distance~$d$. The neutral Coulomb bound four-particle biexciton complex we consider has an interesting charge separation feature with a pair of like charge particles confined to one layer and another pair of the like charge particles of an opposite sign confined to another layer as shown in Fig.~\ref{fig1}~(c).

We use the configuration space method\cite{Bondarev2016} to derive the binding energies for the trion and biexciton complexes in Fig.~\ref{fig1}~(b) and (c). Our model applies both to semiconductor CQWs and to layered TMD heterostructures. The configuration space approach was originally pioneered by Landau,\cite{LandauQM} Gor'kov and Pitaevski,\cite{Pitaevski63} Holstein and Herring\cite{Herring} in their studies of molecular binding and magnetism. The method has been shown to be especially advantageous in the case of quasi-1D semiconductors,\cite{Bondarev11PRB,Bondarev14PRB} where it offers easily tractable, complete analytical solutions to reveal universal asymptotic relations between the binding energy of the complex of interest and the binding energy of the exciton in the same nanostructure. In this approach, the trion or biexciton bound state forms due to the exchange under-barrier tunneling between the \emph{equivalent} configurations of the electron-hole system in the configuration space. For example, in the trion complex of an electron and a direct exciton in Fig.~\ref{fig1}~(a) the exciton configurations ($h-e_1$) and ($h-e_2$) are inequivalent, whereas they are obviously equivalent (can be viewed as two equivalent excitons sharing the same hole) in the trion complex of an electron and an indirect exciton in Fig.~\ref{fig1}~(b). For exciton complexes like that the strength of the binding is controlled by the exchange tunneling rate between the equivalent configurations of the electron-hole system.\cite{Bondarev2016} The binding energy is then given by the tunnel exchange integral determined by means of an appropriate variational procedure. As any variational approach, the method gives an upper bound for the (negative) \emph{ground} state binding energy of the exciton complex of interest. The method captures the essential kinematics of the exciton complex formation and helps understand in simple terms the general physical principles to underlie its stability.

The article is structured as follows.~Section~\ref{sec2} formulates the Hamiltonian, derives the tunnel exchange integral, and obtains the binding energy expression for the biexciton complex formed by two indirect excitons as shown in Fig.~\ref{fig1}~(c). Section~\ref{sec3} uses the results of Sec.~\ref{sec2} to address the simpler case of the trion complex shown in Fig.~\ref{fig1}~(b). Section~\ref{sec4} compares and discusses the results of the previous two sections. Section~\ref{sec5} summarizes and concludes the article.

\section{Biexciton complex formed by two indirect excitons}\label{sec2}

This four-particle problem is initially formulated for two interacting ground-state indirect excitons sketched in Fig~\ref{fig1}~(c). One can easily see that the problem is effectively one-dimensional in view of the fact that the lowest total energy of the four-particle system of two indirect excitons is achieved when the in-plane projections $\rho_{1,2}$ of the relative electron-hole coordinates $\textbf{r}_{1,2}\!=\!\textbf{r}_{e1,2}-\textbf{r}_{h1,2}$ fall on the straight line connecting the centers of mass of the excitons separated by the distance $\Delta\rho$, see Fig~\ref{fig1}~(c). The intraexciton motion can be legitimately treated as being much faster than the interexciton center-of-mass relative motion since the exciton itself is normally more stable than any of its compound complexes. Therefore, the adiabatic approximation can be used to simplify the formulation of the problem by separating out intra- and interexciton motion coordinates. With no substantial loss of generality one can also treat the parallel layers of semiconductor material as being of zero thickness. Such a model assumption can be justified by the obvious fact that due to the electron-hole Coulomb attraction in indirect excitons their ground-state energy can only be the lowest when the electron and hole single-particle wave functions are centered on the inner walls of the bilayer system, regardless of how thick the actual material layers are (CQWs, TMDs, or bilayer graphene). With this in mind, one can assign the two \emph{independent} in-plane projections $\rho_1$ and $\rho_2$ of the relative electron-hole coordinates $\textbf{r}_1$ and $\textbf{r}_2$ to represent the effective configuration space ($\rho_1$, $\rho_2$), in which the ground-state Hamiltonian of the two \emph{equivalent} interacting indirect excitons [sketched in Fig.~\ref{fig1}~(c)] takes the following form
\begin{widetext}
\begin{eqnarray}
\hat{H}(\rho_1,\rho_2,\Delta\rho,d)=-\frac{1}{\rho_1}\frac{\partial}{\partial\,\!\rho_{1}}\,\rho_1\frac{\partial}{\partial\,\!\rho_{1}}
-\frac{1}{\rho_2}\frac{\partial}{\partial\,\!\rho_{2}}\,\rho_2\frac{\partial}{\partial\,\!\rho_{2}}\hskip1.5cm\label{biexcham}\\
-\frac{1}{\sqrt{\rho_1^2+d^2}}-\frac{1}{\sqrt{\rho_2^2+d^2}}-\frac{1}{\sqrt{(\rho_1-\Delta\rho)^2+d^2}}-\frac{1}{\sqrt{(\rho_2+\Delta\rho)^2+d^2}}\nonumber\\
-\frac{2}{\sqrt{[(\sigma\rho_1+\rho_2)/\lambda+\Delta\rho]^2+d^2}}-\frac{2}{\sqrt{[(\rho_1+\sigma\rho_2)/\lambda-\Delta\rho]^2+d^2}}\hskip0.4cm\nonumber\\
+\frac{2}{|\sigma(\rho_1-\rho_2)/\lambda+\Delta\rho|}+\frac{2}{|(\rho_1-\rho_2)/\lambda-\Delta\rho|}\;.\hskip1.5cm\nonumber
\end{eqnarray}
\end{widetext}
Here, the "atomic units"\space are used,\cite{Bondarev2016,LandauQM,Pitaevski63,Herring} whereby the distance and the energy are measured in units of the exciton Bohr radius $a^\ast_B\!=\!0.529\,\mbox{\AA}\,\varepsilon/\mu$ and the exciton Rydberg energy $Ry^\ast\!=\hbar^2/(2\mu\,m_0a_B^{\ast2})\!=\!13.6\,\mbox{eV}\,\mu/\varepsilon^2$, respectively. The parameters $\mu\!=\!m_e/(\lambda\,m_0)$ and $\sigma\!=m_e/m_h$ stand for the exciton reduced effective mass (in units of the free electron mass $m_0$) and the electron-to-hole effective mass ratio, respectively, $\lambda\!=\!1+\sigma$, the constant $\varepsilon$ represents the \emph{effective} average dielectric constant for the bilayer structure, and the image-charge effects are neglected.\cite{LeavittLittle} The first two lines in Eq.~(\ref{biexcham}) describe two non-interacting 1D indirect excitons. Their individual Coulomb potentials are symmetrized to account for the presence of the neighbor a distance~$\Delta\rho$ away, as seen from the $\rho_1$- and $\rho_2$-coordinate systems treated independently [see Fig.~\ref{fig1}~(c)]. The last two lines are the interexciton exchange Coulomb interactions --- electron-hole (line next to last) and hole-hole + electron-electron (last line), respectively.~The strong transverse confinement in reduced dimensionality semiconductors is known to result in the mass reversal effect,\cite{HaugKoch,Cardona} whereby the bulk heavy hole state (the one that forms the \emph{lowest} excitation energy exciton) acquires a longitudinal mass comparable to the bulk \emph{light} hole mass ($\approx\!m_e$), to result in $m_h\!\approx m_e$ in our case. So we keep $\sigma\!=\!1$ in Eq.~(\ref{biexcham}) for simplicity in what follows, which thereby provides the \emph{upper} bounds for the (negative) binding energies of the exciton complexes of interest (meaning that the experimental binding energies in systems of relevance can possibly exceed but cannot be smaller then those we are about to derive).

The biexciton binding energy $E_{X\!X}$ as a function of the interlayer distance $d$ is given by $E_{X\!X}(d)\!=\!E_g-2E_{I\!X}(d),$ where $E_g$ is the lowest eigenvalue of the Hamiltonian~(\ref{biexcham}) and $E_{I\!X}$ is the binding energy of the ground-state indirect exciton. Negative $E_{X\!X}$ indicates that the biexciton is stable with respect to the dissociation into two isolated indirect excitons. The eigenvalue problem for the indirect exciton was earlier studied by Leavitt and Little.\cite{LeavittLittle} Their results we use here are as follows (atomic units)
\begin{equation}
E_{I\!X}(d)=\alpha^2-\frac{4\alpha+4\alpha^4d^2E_1(2\alpha d)\exp(2\alpha d)}{1+2\alpha d}\,.
\label{IXEn}
\end{equation}
Here, $E_1(x)\!=\!\int_{x}^{\infty}\!dt\,e^{-t}\!/t$ is the exponential integral and
\begin{equation}
\alpha=\frac{2}{1+2\sqrt{d}}\;.
\label{alpha}
\end{equation}
The corresponding wave function of the \emph{in-plane} relative electron-hole motion in the ground-state indirect exciton is of the following form
\begin{equation}
\psi_{I\!X}(\rho,d)=N\exp[-\alpha(\sqrt{\rho^2+d^2}-d)],
\label{IXwfunc}
\end{equation}
with the normalization constant
\begin{equation}
N=\frac{4}{\sqrt{1+4\sqrt{d}+8d\,(1+\sqrt{d}\,)}}
\label{N}
\end{equation}
determined by the condition
\[
\int_{0}^{\infty}\!\!\!d\rho\,\rho\,|\psi_{I\!X}(\rho,d)|^2=1\,.
\]

\begin{figure}[t]
\epsfxsize=8.5cm\centering{\epsfbox{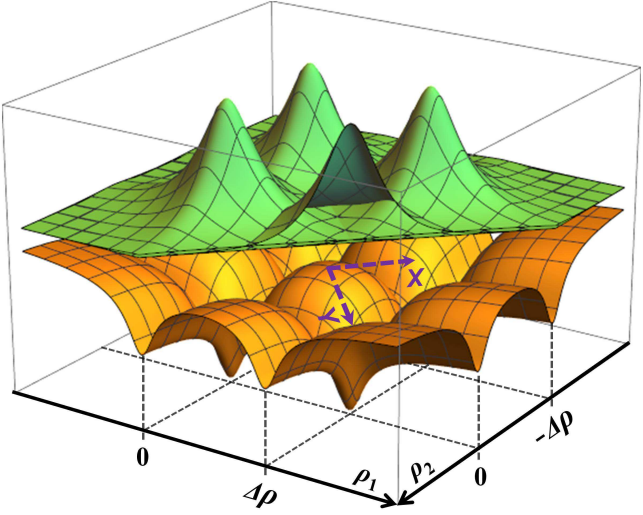}}\caption{(Color online) Schematic of the tunnel exchange coupling configuration for the two ground-state indirect excitons to form the biexciton (or the trion) complex. The coupling occurs in the configuration space of the two \emph{independent} in-plane relative electron-hole motion coordinates, $\rho_1$ and $\rho_2$, of each of the excitons (separated by the center-of-mass-to-center-of-mass distance $\Delta\rho$ --- cf. Fig.~\ref{fig1}), due to the tunneling of the system through the potential barriers formed by the two single-exciton Coulomb interaction potentials (bottom) given by the second line in Eq.~(\ref{biexcham}), between the equivalent states represented by the isolated two-exciton wave functions given by Eqs.~(\ref{psi0xy}) and (\ref{IXwfunc}) and shown on the top.}\label{fig2}
\end{figure}

The Hamiltonian (\ref{biexcham}) is effectively two-dimensional in the configuration space of the two \emph{independent} in-plane projections $\rho_1$ and $\rho_2$ of the relative electron-hole coordinates $\textbf{r}_1$ and $\textbf{r}_2$ of the two indirect excitons. Figure~\ref{fig2}, bottom, shows schematically the potential energy surface of the two closely spaced non-interacting indirect excitons [the second line in Eq.~(\ref{biexcham})] in the configuration space ($\rho_1$, $\rho_2$). The surface has four symmetrical minima to represent isolated \emph{equivalent} two-exciton states (shown in Fig.~\ref{fig2}, top). These minima are separated by the potential barriers responsible for the tunnel exchange coupling between the equivalent states of the system in the configuration space. The coordinate transformation
\begin{equation}
x=\frac{\rho_1-\rho_2-\Delta\rho}{\sqrt{2}}\,,\hskip0.5cm y=\frac{\rho_1+\rho_2}{\sqrt{2}}
\label{transformation}
\end{equation}
places the origin of the new coordinate system ($x$, $y$) into the intersection of the two tunnel channels between the respective potential minima in Fig.~\ref{fig2}, whereby the exchange splitting formula of Refs.~\onlinecite{LandauQM,Pitaevski63,Herring} takes the form
\begin{equation}
E_{g,u}(\Delta\rho)-2E_{I\!X}=\mp J(\Delta\rho).
\label{Egu}
\end{equation}
Here, $E_{g,u}(\Delta\rho)$ are the ground-state and excited-state energies, eigenvalues of the Hamiltonian~(\ref{biexcham}), of the two coupled excitons as functions of their center-of-mass-to-center-of-mass separation distance $\Delta\rho$, and $J(\Delta\rho)$ is the tunnel exchange coupling integral responsible for the bound state formation of two excitons. For the biexciton, this takes the form\cite{Bondarev11PRB,Bondarev2016}
\begin{equation}
J_{X\!X}(\Delta\rho)=\!\frac{2}{3!}\!\int_{\!-\Delta\rho/\!\sqrt{2}}^{\Delta\rho/\!\sqrt{2}}dy\!\left|\psi_{X\!X}(x,y)\frac{\partial\psi_{X\!X}(x,y)}{\partial x}\right|_{x=0}\!\!\!
\label{JXX}
\end{equation}
with $\psi_{X\!X}(x,y)$ being the solution to the Schr\"{o}dinger equ\-a\-tion with the Hamiltonian~(\ref{biexcham}) transformed to the $(x,y)$ coordinates according to Eq.~(\ref{transformation}). The factor $2/3!$ comes from the two equivalent tunnel channels being present in the biexciton problem as one can see in Fig.~\ref{fig2}, to mix up three equivalent indistinguishable two-exciton states in the configuration space --- one state is given by the two minima on the $x$-axis and two more states are represented by each of the minima on the $y$-axis.

The function $\psi_{X\!X}(x,y)$ in Eq.~(\ref{JXX}) is sought in the form
\begin{equation}
\psi_{X\!X}(x,y)=\phi_{I\!X}(x,y)\exp[-S_{X\!X}(x,y)]\,,
\label{psiXXxy}
\end{equation}
where
\begin{equation}
\phi_{I\!X}(x,y)\!=\psi_{I\!X}\!\left[\rho_1(x,y),d\,\right]\psi_{I\!X}\!\left[\rho_2(x,y),d\,\right]
\label{psi0xy}
\end{equation}
is the product of the two single-exciton wave functions (shown in Fig.~\ref{fig2}, top) given by Eq.~(\ref{IXwfunc}) in which $\rho_1$ and $\rho_2$ are expressed in terms of $x$ and $y$ using Eq.~(\ref{transformation}). The function $\phi_{I\!X}(x,y)$ represents the isolated two-exciton state centered at the minimum $\rho_1\!=\!\rho_2\!=\!0$ (or $x\!=\!-\Delta\rho/\sqrt{2}$, $y\!=\!0$) of the configuration space potential in Fig.~\ref{fig2}.~This is the approximate solution to the Shr\"{o}dinger equation with the Hamiltonian given by the first two lines in Eq.~(\ref{biexcham}). The function $S_{X\!X}(x,y)$ in Eq.~(\ref{psiXXxy}) is assumed to be a smooth slowly varying function in the domain of interest --- the square $|x|,|y|\le\Delta\rho/\sqrt{2}$ in Fig.~\ref{fig2} --- to account for the major deviation of $\psi_{X\!X}(x,y)$ from $\phi_{I\!X}(x,y)$ in its "tail area" $x\!\sim\!y\!\sim\!0$ due to the tunnel exchange coupling to another equivalent isolated two-exciton state centered at $\rho_1\!=\Delta\rho$, $\rho_2\!=\!-\Delta\rho$ (or $x\!=\!\Delta\rho/\sqrt{2}$, $y\!=\!0$).

Substituting Eq.~(\ref{psiXXxy}) into the Schr\"{o}dinger equation with the Hamiltonian (\ref{biexcham}) pre-transformed to the $(x,y)$ coordinates, one obtains
\begin{eqnarray}
\left[\frac{1}{\phi_{I\!X}}\frac{\partial\phi_{I\!X}}{\partial x}-\frac{x+\Delta\rho/\sqrt{2}}{y^2-(x+\Delta\rho/\!\sqrt{2})^2}
\right]\frac{\partial S_{X\!X}}{\partial x}\hskip0.5cm\nonumber\\
+\left[\frac{1}{\phi_{I\!X}}\frac{\partial\phi_{I\!X}}{\partial y}+\frac{y}{y^2-(x+\Delta\rho/\sqrt{2})^2}
\right]\frac{\partial S_{X\!X}}{\partial y}\hskip0.5cm\nonumber\\
=\frac{1}{\sqrt{(y/\sqrt{2}+\Delta\rho)^2+d^2}}+\frac{1}{\sqrt{(y/\sqrt{2}-\Delta\rho)^2+d^2}}\nonumber\\
-\frac{1}{|x/\sqrt{2}+3\Delta\rho/2|}-\frac{1}{|x/\sqrt{2}-\Delta\rho/2|}\,,\hskip1cm\label{SPDE}
\end{eqnarray}
where the righthand side is the \emph{half} of the interexciton exchange Coulomb interaction energy given by the last two lines in the Hamiltonian~(\ref{biexcham}). This is valid up to negligible terms of the order of the interexciton van der Waals energy ($\sim\!1/\Delta\rho^6$) and up to the second order derivatives of~$S_{X\!X}$ (a~slowly varying function).\cite{LandauQM,Pitaevski63,Herring} This can be further simplified in view of the fact that we only need $S_{X\!X}(x,y)$ in the neighborhood of the origin of the $(x,y)$ coordinate system to be able to evaluate the tunnel exchange coupling integral~(\ref{JXX}) to an acceptable accuracy. Bearing in mind that $\Delta\rho>1$ since in the stable complex of two excitons the interexciton center-of-mass-to-center-of-mass distance should be greater than the exciton Bohr radius, one has the following series expansions
\begin{eqnarray}
\frac{1}{\phi_{I\!X}}\frac{\partial\phi_{I\!X}}{\partial x}-\frac{x+\Delta\rho/\sqrt{2}}{y^2-(x+\Delta\rho/\!\sqrt{2})^2}
\approx\sqrt{2}\left(\!-\alpha+\frac{1}{\Delta\rho}\right),\nonumber\\
\frac{1}{\phi_{I\!X}}\frac{\partial\phi_{I\!X}}{\partial y}+\frac{y}{y^2-(x+\Delta\rho/\sqrt{2})^2}\approx-\frac{2y}{\Delta\rho^2}\,,\hskip0.75cm\nonumber\\
\frac{1}{\sqrt{(y/\sqrt{2}+\Delta\rho)^2+d^2}}+\frac{1}{\sqrt{(y/\sqrt{2}-\Delta\rho)^2+d^2}}\hskip0.5cm\nonumber\\
\approx\frac{2}{\Delta\rho}+\frac{y^2-d^2}{\Delta\rho^3}+\frac{2y^4-12y^2d^2+3d^4}{4\Delta\rho^5}\hskip1cm\label{seriesexp}
\end{eqnarray}
to the first few orders in small parameter $1/\Delta\rho$, whereby Eq.~(\ref{SPDE}) simplifies to take the form as follows
\begin{eqnarray}
\sqrt{2}\left(\!-\alpha+\frac{1}{\Delta\rho}\right)\frac{\partial S_{X\!X}}{\partial x}\hskip2.0cm\nonumber\\
\approx\frac{2}{\Delta\rho}-\frac{1}{|x/\sqrt{2}+3\Delta\rho/2|}-\frac{1}{|x/\sqrt{2}-\Delta\rho/2|}\hskip.5cm\label{SPDEapp}
\end{eqnarray}
to the \emph{leading} order in $1/\Delta\rho$. This is to be solved with the boundary condition $S_{X\!X}(-\Delta\rho/\sqrt{2},y)\!=\!0$ coming from the fact that $\psi_{X\!X}(-\Delta\rho/\sqrt{2},y)\!=\phi_{I\!X}(-\Delta\rho/\sqrt{2},y)$. In the domain where $|x|,|y|\!\le\!\Delta\rho/\sqrt{2}$ we are interested in (see Fig.~\ref{fig2}), one obtains
\begin{eqnarray}
S_{X\!X}(x,y)\!=\!\frac{\Delta\rho}{\alpha\Delta\rho-\!1}\!\left(\ln\!\left|\frac{x\!+\!3\Delta\rho/\!\sqrt{2}}{x\!-\!\Delta\rho/\!\sqrt{2}}\right|
\!-\!\frac{\sqrt{2}x}{\Delta\rho}-1\!\right)\!\!.\hskip0.5cm
\label{sXXxy}
\end{eqnarray}
This function is only an approximation of the actual function $S_{X\!X}(x,y)$ defined by the partial differential equation in Eq.~(\ref{SPDE}). However, this approximation is good enough in the main domain of our interest here --- in the neighborhood of the origin of the $(x,y)$ coordinate system in Fig.~\ref{fig2} where the main tunnel probability flow occurs.

The function one obtains by plugging Eq.~(\ref{sXXxy}) into Eq.~(\ref{psiXXxy}) can be used to evaluate the tunnel exchange coupling integral in Eq.~(\ref{JXX}).~In so doing, it is legitimate to neglect a very weak $y$-dependence of the integrand since it was neglected in Eqs.~(\ref{SPDEapp}) and (\ref{sXXxy}). With this one has
\begin{eqnarray}
J_{X\!X}(\Delta\rho)=\frac{2N^4}{3}\Delta\rho\!\left(\!\frac{\alpha\Delta\rho}{\sqrt{\Delta\rho^2+4d^2}}+\frac{1}{3(\alpha\Delta\rho-1)}\!\right)\hskip0.3cm\nonumber\\
\times\!\left(\frac{e}{3}\right)^{2\Delta\rho/(\alpha\Delta\rho-1)}\!\exp\!\left[-2\alpha\!\left(\!\sqrt{\Delta\rho^2+4d^2}-2d\right)\right]\!.\hskip0.5cm
\label{JXXfin}
\end{eqnarray}
The structure of this expression suggests that it has the maximum at some $\Delta\rho=\Delta\rho_0$. Indeed, it tends to become a negative when $\alpha\Delta\rho<1$ in the second term in the parentheses in front of the exponential factor. This will always be the case when $d$ is large enough to make this term $-1/3$ with $\alpha\!=\!2/(1+2\sqrt{d})\!\approx\!1/\sqrt{d}\!\sim\!0$, whereby the first term in the parentheses becomes negligible. For $\alpha\Delta\rho>1$, on the other hand, this expression is seen to be manifestly positive, approaching zero as $\Delta\rho$ increases. Consequently, the energy $E_g(\Delta\rho)$ in Eq.~(\ref{Egu}) will have the negative minimum (biexcitonic state) at $\Delta\rho=\Delta\rho_0$. Extremum seeking for the function $J_{X\!X}(\Delta\rho)$ in Eq.~(\ref{JXXfin}), subject to the condition $\Delta\rho>1$ in order to \emph{only} include the leading terms in $1/\Delta\rho$ for consistency with Eqs.~(\ref{SPDEapp}) and (\ref{sXXxy}), results in
\begin{equation}
\Delta\rho_0^{\mbox{\tiny\emph{XX}}}=\frac{7\alpha-2/3}{2\alpha^2}\,.
\label{Drho0XX}
\end{equation}

One last important step to add to our analysis is to take into account the higher order expansion terms of the interexciton long-range interaction energy \emph{at a distance} such as the dipolar (DI) and quadrupolar (QI) interaction contributions --- which are greater than the interexciton van der Walls interaction energy we neglected in the main equation~(\ref{SPDE}), which are present in the square-root series expansion in Eq.~(\ref{seriesexp}), and which we did miss out on using the leading term approximation in $1/\Delta\rho$ in Eqs.~(\ref{SPDEapp}) and (\ref{sXXxy}). We complete this step by including these terms as given by Eq.~(\ref{seriesexp}) taken at $y=0$, to write down the final expression for the biexciton binding energy as follows
\begin{equation}
E_{X\!X}=-J_{X\!X}(\Delta\rho_0^{\mbox{\tiny\emph{XX}}})+\frac{2d^2}{(\Delta\rho_0^{\mbox{\tiny\emph{XX}}})^3}-\frac{3d^2}{2(\Delta\rho_0^{\mbox{\tiny\emph{XX}}})^5}\,.
\label{EXX}
\end{equation}

Equations~(\ref{JXXfin})--(\ref{EXX}) and (\ref{alpha}) solve the ground-state binding energy problem for the biexciton complex formed by two indirect dipolar excitons in layered quasi-2D nanostructures.

\section{Charge-coupled indirect exciton: the trion}\label{sec3}

The trion system we study here is a charged three-particle complex formed by an indirect exciton and a hole (electron). In such a complex, two like charge carriers are confined to the \emph{same} layer and the third (opposite charge) carrier is confined to another layer. The system can be viewed as two like charge carriers sharing the third carrier of an opposite sign to form two \emph{equivalent} indirect exciton configurations (as shown Fig.~\ref{fig1}~(b) for the negative trion complex of the two electrons sharing the same hole). The binding energy can then be found using a modification of the Hamiltonian (\ref{biexcham}), in which the first two lines are the same as in Eq.~(\ref{biexcham}), the line next to last is absent, and only one term is present in the last line --- either the first or the second one for the positive and negative trion complex, respectively. With $\sigma\!=\!1$ we have chosen to use in the Hamiltonian for the reasons above, the positive-negative trion binding energy difference disappears, and the quantity we are about to derive will provide the upper bound for the (negative) ground-state binding energy of the trion complex of interest.

Just like in the case of the biexciton above, the treatment of the trion problem starts with the coordinate transformation~(\ref{transformation}) to bring the Hamiltonian from the original configuration space ($\rho_1,\rho_2$) into the coordinate space $(x,y)$, whereby the origin and both coordinate axes can be adjusted as shown in Fig.~\ref{fig2} to capture the maximal tunnel flow that occurs at the intersection of the two tunnel channels between the respective minima of the potential energy surface. The tunnel exchange splitting integral in Eq.~(\ref{Egu}) now takes the form\cite{Bondarev14PRB,Bondarev2016}
\begin{equation}
J_{X^{^{\!\ast}}}(\Delta\rho)=\int_{\!-\Delta\rho/\!\sqrt{2}}^{\Delta\rho/\!\sqrt{2}}\!dy\!\left|\psi_{X^{^{\!\ast}}}(x,y)\frac{\partial\psi_{X^{^{\!\ast}}}(x,y)}{\partial x}\right|_{x=0},
\label{JXast}
\end{equation}
where $\psi_{X^{^{\!\ast}}}(x,y)$ is the ground-state wave function of the Schr\"{o}dinger equation with the Hamiltonian (\ref{biexcham}) modified to the negative trion case and transformed to the $(x,y)$ coordinates as described above. [We do the negative trion both for definiteness and for simplicity since the positive trion case would require a counterpart of the transformation~(\ref{transformation}) to be used to bring the first term in the last line of the Hamiltonian~(\ref{biexcham}) to the form one obtains using Eq.~(\ref{transformation}) for the second term in the negative trion case.] The tunnel exchange current integral $J_{X^{^{\ast}}}(\Delta\rho)$ is due to the electron position exchange relative to the hole --- see Fig.~\ref{fig1}~(b). This corresponds to the tunneling of the entire three particle system between the two equivalent indistinguishable configurations of the two indirect excitons sharing the same hole in the configuration space $(\rho_1,\rho_2)$. They are those given by the minima at $\rho_1\!=\!\rho_2\!=\!0$ and $\rho_1\!=\!-\rho_2\!=\!\Delta\rho$ in Fig.~\ref{fig2}. Such a tunnel exchange binds the three particle system to form a stable trion state.

Like in the case of the biexciton, one seeks the function $\psi_{X^{^{\!\ast}}}(x,y)$ in the form
\begin{equation}
\psi_{X^{^{\!\ast}}}(x,y)=\phi_{I\!X}(x,y)\exp[-S_{X^{^{\!\ast}}}(x,y)]
\label{psixy}
\end{equation}
with $\phi_{I\!X}$ given by Eqs.~(\ref{psi0xy}) and (\ref{IXwfunc}), and $S_{X^{^{\!\ast}}}$ representing a smooth \emph{slowly} varying function to account for the deviation of $\psi_{X^{^{\!\ast}}}$ from $\phi_{I\!X}$ in the "tail area" ($x\!\sim\!y\!\sim\!0$) due to the tunnel exchange coupling of the isolated two-exciton state centered at $\rho_1\!=\!\rho_2\!=\!0$ ($x\!=\!-\Delta\rho/\sqrt{2}$, $y\!=\!0$) to another equivalent isolated two-exciton state centered at $\rho_1\!=\!-\rho_2\!=\!\Delta\rho$ ($x\!=\!\Delta\rho/\sqrt{2}$, $y\!=\!0$) as shown in~Fig.~\ref{fig2}. Substituting Eq.~(\ref{psixy}) into the Schr\"{o}dinger equation with the negative trion Hamiltonian pre-transformed to the $(x,y)$ coordinates and following exactly the steps described above for the biexciton case, one obtains
\begin{equation}
\sqrt{2}\left(\!-\alpha+\frac{1}{\Delta\rho}\right)\frac{\partial S_{X^{^{\!\ast}}}}{\partial x}\approx-\frac{1}{|x/\sqrt{2}-\Delta\rho/2|}
\label{SPDEapp1}
\end{equation}
with the solution of the form
\begin{equation}
S_{X^{^{\!\ast}}}(x,y)\!=\!\frac{\Delta\rho}{\alpha\Delta\rho-\!1}\ln\!\left|\frac{\!\sqrt{2}\Delta\rho}{x\!-\!\Delta\rho/\!\sqrt{2}}\right|
\label{sXxy}
\end{equation}
to fulfill the boundary condition $S_{X^{^{\!\ast}}}(-\Delta\rho/\sqrt{2},y)\!=\!0$ in the domain of interest $|x|,|y|\!\le\!\Delta\rho/\sqrt{2}$. Using this in the tunnel exchange coupling integral of Eq.~(\ref{JXast}), by analogy with Eq.~(\ref{JXXfin}) one now obtains
\begin{eqnarray}
J_{X^{^{\!\ast}}}(\Delta\rho)=2N^4\Delta\rho\!\left(\!\frac{\alpha\Delta\rho}{\sqrt{\Delta\rho^2+4d^2}}+\frac{1}{\alpha\Delta\rho-1}\!\right)\hskip0.5cm\nonumber\\
\times\!\left(\frac{1}{2}\right)^{\!2\Delta\rho/(\alpha\Delta\rho-1)}\!\exp\!\left[-2\alpha\!\left(\!\sqrt{\Delta\rho^2+4d^2}-2d\right)\right]\!.\hskip0.5cm
\label{JXfin}
\end{eqnarray}
This expression has the same properties as its biexciton counterpart in Eq.~(\ref{JXXfin}) does, with the only difference being that now the same extremum seeking procedure applied to the function $J_{X^{^{\!\ast}}}(\Delta\rho)$ results in
\begin{equation}
\Delta\rho_0^{\mbox{\tiny\emph{X}}^{\!\ast}}=\frac{7\alpha-2}{2\alpha^2}\,,
\label{Drho0Xast}
\end{equation}
to give the trion binding energy expression in the form
\begin{equation}
E_{X^{^{\!\ast}}}=-J_{X^{^{\!\ast}}}(\Delta\rho_0^{\mbox{\tiny\emph{X}}^{\!\ast}})\,.
\label{EXast}
\end{equation}

\begin{figure}[t]
\epsfxsize=8.5cm\centering{\epsfbox{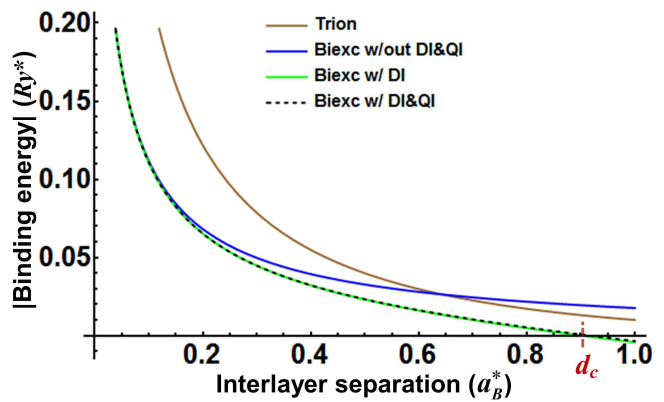}}\caption{(Color online) Binding energies of the biexciton and trion complexes as functions of the interlayer separation distance, calculated in atomic units according to Eqs.~(\ref{JXXfin})--(\ref{EXX}) and Eqs.~(\ref{JXfin})--(\ref{EXast}), respectively. Biexciton binding energy is presented by the three different curves for comparison, calculated with no long-range interaction terms included, with only the dipolar interaction term included, and with both dipolar and quadrupolar interaction terms included, respectively.}\label{fig3}
\end{figure}

\section{Discussion}\label{sec4}

Figure~\ref{fig3} shows the binding energies in atomic units for the biexciton and trion complexes calculated according to Eqs.~(\ref{JXXfin})--(\ref{EXX}) and Eqs.~(\ref{JXfin})--(\ref{EXast}), respectively, as functions of the interlayer separation distance. Since the long-range DI and QI terms in Eq.~(\ref{EXX}) have the opposite signs, being repulsive and attractive, respectively, the biexciton binding energy is presented by three curves calculated with no DI and QI terms included, with only the DI term included, and with both DI and QI terms included, respectively. The QI term is seen not to contribute much at all. Therefore, we only take into account the DI term in Eq.~(\ref{EXX}) in the discussion of our results in what follows.

Figure~\ref{fig3} shows the presence of the critical \emph{threshold} interlayer separation $d_c\!\approx\!0.9$ for the biexciton complex --- the cutoff distance with the biexciton formation being only possible at $d\!<\!d_c$ due to the long-range dipolar repulsion of the two indirect excitons --- in agreement with what was earlier reported by Meyertholen and Fogler.\cite{Fogler08} A similar repulsive interaction is absent from the trion complex, and so no critical distance exists for the trion and its binding energy goes down to zero exponentially always exceeding that of the biexciton as the interlayer separation increases.

\begin{figure}[t]
\epsfxsize=8.5cm\centering{\epsfbox{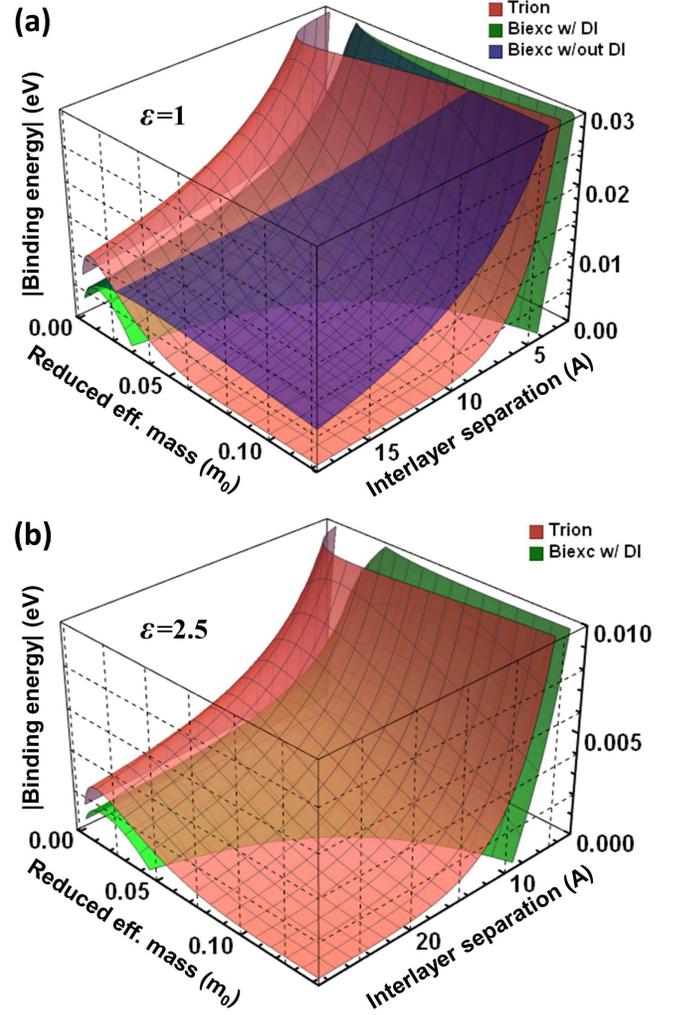}}\caption{(Color online) Binding energies of the biexciton and trion complexes as functions of the interlayer separation distance and the exciton reduced effective mass, calculated in absolute units according to Eqs.~(\ref{JXXfin})--(\ref{EXX}) and Eqs.~(\ref{JXfin})--(\ref{EXast}), respectively, for $\varepsilon\!=\!1$~(a) and $\varepsilon\!=\!2.5$~(b). In~(a), the biexciton binding energy is presented by the two different graphs calculated with no long-range interaction terms included and with the dipolar interaction term included, respectively, for the purpose of comparison.}\label{fig4}
\end{figure}

As it follows from our original model Hamiltonian~(\ref{biexcham}), the binding energies in Eqs.~(\ref{EXX}) and (\ref{EXast}) are generally functions of the interlayer distance $d$, the exciton reduced effective mass $\mu$, and the \emph{effective} average dielectric constant $\varepsilon$ for the bilayer structure. Figure~\ref{fig4} shows the biexciton and trion binding energies calculated in absolute units as functions of $d$ and $\mu$ with $\varepsilon\!=\!1$ in (a) and $\varepsilon\!=\!2.5$ in (b) for the purpose of comparison.~Both of the dielectric constants we have chosen are representative of quasi-2D layered van der Waals materials such as bilayer graphene and quasimonolayer TMD systems where the material layers are generally surrounded by air (or by dielectrics with relatively small permittivity). Hence, the value $\varepsilon\!=\!1$ is expected to represent reasonably well the interlayer electron-hole Coulomb interaction in \emph{indirect} excitons, whereas $\varepsilon\!=\!2.5$ corresponds precisely to $\varepsilon_{ef\!f}\!=\!5$ reported in Ref.~\onlinecite{Wang17} as being the effective dielectric screening parameter to give a realistic binding energy estimate $\sim\!0.5$~eV for \emph{direct} 2D excitons in TMD monolayers (with $\mu\!=\!0.25m_0$ and the \emph{four-fold} increased Rydberg constant typical of 2D confinement as opposed to the unchanged atomic Rydberg of $13.6$~eV we use here).

In Fig.~\ref{fig4}~(a), two different graphs are presented for the biexciton binding energy --- calculated with and with no long-range DI term included in Eq.~(\ref{EXX}), respectively --- to compare with that for the trion. One can see that taking account of the long-range dipolar repulsion in the system of two indirect excitons is important to obtain the correct estimate for the biexciton binding energy. With no DI term included one obtains an illusive crossover behavior for the biexciton and trion binding energies, whereby being less for relatively small $\mu$ and $d$ the biexciton binding energy increases to become greater than that of the trion with $\mu$ and $d$ increasing. This is quite similar to what was reported earlier for quasi-1D biexciton and trion systems.\cite{Bondarev14PRB} For compound quasi-1D exciton systems this is a generic feature resulted from the fact that greater $\mu$, while not affecting significantly the single charge tunnel exchange in the trion complex, make the neutral biexciton complex generally more compact to facilitate the mixed charge tunnel exchange in it and thus to make the biexciton binding energy greater than that of the trion. This feature is suppressed in quasi-2D complexes of \emph{indirect} excitons due to their long-range dipolar repulsion, which is absent from the trion complex to keep its binding energy always being greater than that of the biexciton as one can see in Fig.~\ref{fig4}~(a). One can also see that the biexciton formation cutoff distance $d_c$ is a rapidly decreasing function of $\mu$ --- obviously due to the fact that the biexciton complex tends to get more compact with increasing $\mu$, thereby loosing its ability to withstand the interlayer distance increase.

From Fig.~\ref{fig4}~(b) one can see that the features just described are preserved for $\varepsilon\!=\!2.5$ as well, although the increased dielectric screening results in the overall binding energy decrease both for the biexciton and for the trion to roughly the same extent. The binding energies of both complexes are seen to be only weakly $\mu$ dependent while exceeding $10$~meV and a few tens of meV for the biexciton and trion, respectively, at interlayer separations $d\!\lesssim\!5$~\AA ~typical of layered van der Waals heterostructures. For $d$ greater than that the trion graphs both in (b) panel and in (a) panel of Fig.~\ref{fig4} suggest that the formation of trions can still be quite possible at low temperatures in layered semiconductor materials with relatively small $\mu\!\sim\!(0.02-0.06)m_0$ such as CQWs of some of conventional group-IV/III-V/II-VI semiconductors,\cite{Cardona} whereas the biexciton formation is hardly possible with no extra lateral confinement.\cite{Govorov13,Schinner2013} Comparing~(b) and (a) panels, one can also see that the enhanced screening leads to the increase of the biexciton formation cutoff distance $d_c$, especially for greater $\mu$. The reason is that the Bohr radius increase with $\varepsilon$, on the one hand, and the Rydberg constant increase with $\mu$, on the other, make the exciton larger in size while still keeping it stable enough to be able to form biexciton complexes at larger interlayer separation distances.

Recently,\cite{Kezer14} the problem of the trion complex forma\-tion in conventional GaAs/AlGaAs semiconductor CQWs was studied theoretically in great detail for trions composed of a \emph{direct} exciton and an electron (or a hole) located in the neighboring quantum well as sketched above in Fig.~\ref{fig1}~(a). Significant binding energies were predicted on the order of $10$~meV at the interwell separations $d\sim\!10-20$~nm for the lowest energy positive and negative trion states, allowing the authors to propose a possibility for the trion Wigner crystallization phenomenon. For \emph{laterally} confined semi\-conductor CQW heterostructures, the experimental evidence for the controllable formation of the multiexciton Wigner-like molecular complexes of \emph{indirect} excitons (single exciton, biexciton, triexciton, etc.) was reported recently as well.\cite{Govorov13} Significant binding energies (over 10~meV for $d\!\lesssim\!5$~\AA) for the biexciton and trion complexes formed by \emph{indirect} excitons [sketched in Fig.~\ref{fig1}~(b) and (c)] we report about here suggest that this strongly correlated multi\-exciton phenomenon can also be realized in the quasimonolayer van der Waals hetero\-structures such as double bilayer graphene and few-layer TMD systems.\cite{Li2017,Fogler14,Calman2017,Ross2017,Rivera2015,Baranowski2017}

The overall stability of strongly correlated multi\-exciton structures depends on the stability of their elementary components, their building blocks --- trions and biexcitons. Understanding of the basic principles of the trion and biexciton formation is therefore important to understand the nature of the formation of strongly correlated many-particle electron-hole states that are of direct relevance to important fundamental physical phenomena such as the said Wigner crystallization, exciton BEC and superfluidity. For example, one can imagine a coupled charge-neutral "zigzag" shaped electron-hole structure formed by two trions shown in Fig.~\ref{fig1}~(b), one positively and one negatively charged. This six-particle Wigner-like structure will be electrically neutral with positive and negative charge carriers separated in different layers, and it will have the total spin of zero with \emph{non-zero} alignment of single-particle spins in its ground state. Such a correlated electron-hole complex can also be viewed as a triexciton --- a coupled state of the three indirect singlet excitons. Adding one more charge carrier (electron or hole) to this correlated electron-hole state turns it into the state of non-zero net charge and spin with other features still preserved, to allow precise electro- and magnetostatic control and manipulation by its optical and spin properties. This opens up new routes for nonlinear optics and spinoptronics applications with indirect excitons in quasi-2D semiconductor heterostructures.

\section{Conclusion}\label{sec5}

We consider trion and biexciton states formed by indirect (dipolar) excitons in layered quasi-2D semiconductor heterostructures. The trion state we are looking at is a charged three-particle Coulomb bound electron-hole complex with two like charge particles confined to the same layer and the third (opposite sign) charge particle confined to another layer. The charge-neutral biexciton state we deal with is a Coulomb bound four-particle electron-hole complex with an interesting charge separation feature where a pair of like charges is confined to one layer and another pair of like charges of an opposite sign is confined to another layer. We use the configuration space method developed earlier by one us for quasi-1D excitonic systems,\cite{Bondarev2016} to obtain analytical expressions for the binding energies of the biexciton and trion complexes as functions of the interlayer separation distance, the exciton reduced effective mass, and the effective dielectric constant of the system. The method captures the essential kinematics of the complex formation to reveal that, despite a rapid decrease with distance, the binding energies can exceed $10$~meV and a few tens of meV for the biexciton and trion, respectively, for the interlayer distances $\sim\!3\!-\!5$~\AA ~typical of van der Waals heterostructures. Indirect excitons, biexcitons, and trions formed by indirect excitons control the formation of strongly correlated (Wigner-like) electron-hole crystal structures. Significant binding energies we predict herewith suggest that this strongly correlated multiexciton phenomenon of Wigner crystallization can be realized in layered van der Waals heterostructures such as double bilayer graphene and few-layer TMD systems, to open up new routes for nonlinear coherent optical control and spinoptronics applications with indirect excitons.

\section*{Acknowledgments}

I.V.B. is supported by the DOE grant DE-SC0007117. M.R.V. is supported by the CNRS OBELIX grant ANR-15-CE30-0020-02. Discussions with M.Fogler (UCSD), O.Berman and R.Keze\-ra\-shvili (NY$\,\!$CityTech) are gratefully acknowledged.

\end{document}